\documentclass[aps,prd,twocolumn,superscriptaddress,showpacs]{revtex4}
\usepackage{feynmp,amsmath,amssymb,graphicx,subfigure,color,hyperref,bm}

\begin{document}

\title{Plasmon mode as a detection of the chiral anomaly in Weyl semimetals}

\author{Jianhui Zhou}
\email{jhzhou@andrew.cmu.edu}
\affiliation{Department of Physics, Carnegie Mellon University, Pittsburgh, Pennsylvania 15213, USA}

\author{Hao-Ran Chang}
\email{hrchang@mail.ustc.edu.cn}
\affiliation{Department of Physics and Institute of Solid State Physics, Sichuan Normal University, Chengdu, Sichuan 610066, China}

\author{Di Xiao}
\affiliation{Department of Physics, Carnegie Mellon University, Pittsburgh, Pennsylvania 15213, USA}

\begin{abstract}
Weyl semimetals are one kind of three-dimensional gapless semimetal
with nontrivial topology in the momentum space. The chiral anomaly in
Weyl semimetals manifests as a charge imbalance between the Weyl nodes
of opposite chiralities induced by parallel electric and magnetic fields.
We investigate the chiral anomaly effect on the plasmon mode in both
intrinsic and doped Weyl semimetals within the random phase approximation.
We prove that the chiral anomaly gives rise to a different plasmon mode in intrinsic
Weyl semimetals. We also find the chiral anomaly leads to some exotic
properties in the plasmon dispersion in doped Weyl semimetals. Consequently,
the unconventional plasmon mode acts as a signature of the chiral anomaly in
Weyl semimetals, by which the spectrum of plasmon provides a proper way to
detect the Lifshitz transition.
\end{abstract}


\maketitle

\section{Introduction}

Weyl semimetals \cite{Volovik} (SMs) are a new class of gapless topological phase,
which can be seen as three-dimensional (3D) analogs of graphene. Weyl fermions emerge
from the band degenerate points---the Weyl nodes---in the momentum space, which are
characterized by their chirality. Due to the fermion doubling theorem \cite{NielsenNinomiya}, Weyl nodes with opposite chirality always appear in pairs. Each Weyl node behaves
as a magnetic monopole in the momentum space, which acts as the source/drain for
the Berry curvature field~\cite{DXiao}. It has been predicted that this nontrivial
momentum-space topology of Weyl nodes gives rise to a number of novel electromagnetic
responses\cite{HosurPRL,ZyuzinBurkov,Aji,Grushin,JHZhou,SonSpivak,KYYang,Goswami,Ashby,
Landsteiner,Vazifeh,basar,Hughes}. On the material side, Weyl SMs have been proposed
for strongly correlated iridates~\cite{XGWan}, semiconductor heterostructures \cite{BurkovBalents,Halasz}, and other materials \cite{GXuPRL,ZWuBurkov,Krempa,manes}.
In addition, 3D Dirac materials have recently been realized in both Cd$_{3}$As$_{2}$ \cite{Cava,ZJWCA} and Na$_{3}$Bi \cite{YLChen,ZJWNB}, which could greatly facilitate
the search for Weyl SMs.

A remarkable phenomenon associated with Weyl nodes is the so-called chiral anomaly \cite{Adler,BellJackiw}, in which the application of a pair of parallel electric field $\mathbf{E}$ and magnetic field $\mathbf{B}$ induces a charge imbalance between the
two Weyl nodes with opposite chirality. This chiral anomaly can be utilized to detect
3D Weyl SMs in experiments. For example, a large longitudinal magnetoconductivity was
proposed as a consequence of the chiral anomaly \cite{NielsenNinomiya}, which, however,
is difficult to identify unambiguously in magnetotransport data \cite{KimPRL}. Recently,
nonlocal transport \cite{Parameswaran}, optical conductivity \cite{HosurQi,GoswamiOptics},
and optical absorption \cite{AshbyABJ} measurements have also been proposed to probe the
chiral anomaly in 3D Weyl SMs.

In this paper, we propose an alternative detection method of the chiral anomaly
by employing the plasmon mode in 3D Weyl SMs. We show that the chiral anomaly
would lead to a different plasmon mode in intrinsic Weyl SMs. The chiral anomaly causes
a redshift of the frequency of plasmon mode in doped Weyl SMs. Once the small Fermi
surface crosses the Weyl node that corresponds to the Lifshitz transition (LT) point,
the frequency turns out to be a violetshift. Therefore, the plasmon mode can be
regarded as a signature of the chiral anomaly in 3D Weyl SMs. We also show how to
extract the information of the LT point from the plasmon dispersion.

The rest of this paper is organized as follows. In Sec. \ref{Sec:model}, we discuss
the chiral anomaly effect in Weyl SMs and outline the formalism for plasmon. In
Sec. \ref{Sec:undoped}, we prove the existence of a different plasmon mode due to the chiral
anomaly in undoped Weyl SMs and calculate the plasmon dispersion. In Sec. \ref{Sec:doped},
we consider the chiral anomaly effect on the plasmon mode in doped Weyl SMs and discuss
the chiral anomaly-driven Lifshitz transition. In Sec. \ref{Sec:CONCLUSIONS}, we summarize
the main results of this paper. Finally, in Appendixes \ref{Sec:AppendixA} and \ref{Sec:AppendixB}
we give details of the calculation of the free polarization function.

\section{Model and formalism\label{Sec:model}}

We begin with a low-energy effective Hamiltonian for Weyl fermions in the vicinity of
the Weyl node of chirality $\chi=\pm$,
\begin{eqnarray}
\mathcal{H}=\chi\hbar v_{F}\bf{k}\cdot\bm{\sigma}-\mu_{\chi},\label{Hamiltonian}
\end{eqnarray}
where $v_{F}$ is the Fermi velocity, $\bm{\sigma}=\left(\sigma_{x},\sigma_{y},\sigma_{z}\right)$ refers to the three Pauli matrices, and $\mu_{\chi}$ stands for the chirality-dependent
chemical potential given by a superposition of the equilibrium carrier density and the
pumped carrier density originating from the chiral anomaly. The latter grows linearly
with time, but the large momentum internode scattering would counteract this imbalance
of carriers between two Weyl nodes. Eventually the system reaches a nonequilibrium steady
state characterized by an internode relaxation time $\tau_{v}$ that had been evaluated microscopically~\cite{Parameswaran}. Consequently, the density of electrons pumped into
or out of the neighborhood of Weyl node $\chi$ can be expressed as \cite{NielsenNinomiya}
\begin{eqnarray}
\Delta\rho_{\chi}\equiv\chi\frac{e^{2}}{4\pi^{2}\hbar^{2}}{\bf E}\cdot{\bf B}\tau_{v}.\label{rho}
\end{eqnarray}
We also define several chirality-dependent quantities: the Fermi wave vector $k_{F_{\chi}}^{3}=6\pi^{2}n_{\chi}$, the chemical potential $\mu_{\chi}=\hbar v_{F}k_{F\chi}$, and the charge density $n_{\chi}=n+\Delta\rho_{\chi}$. When the two Weyl nodes are equally populated, the corresponding Fermi wave vector and the chemical potential become $k_{F}^{3}=6\pi^{2}n$, $\mu=\hbar v_{F} k_{F}$. For convenience, we restrict our discussion
to $\mathbf{E\cdot B}>0$. For the undoped case with vanishing equilibrium chemical potential,
$\mu_{\chi}$ depends only on the pumped charge associated with the chiral anomaly,
\begin{eqnarray}
\mu_{\chi}=\chi\big(\frac{3e^2\hbar v_{F}^3}{2}
{\bf E}\cdot{\bf B}\tau_{v}\big)^{\frac{1}{3}}.
\label{mupmud}
\end{eqnarray}
Meanwhile, for the doped case with a finite chemical potential $\mu$, we obtain the
corresponding chirality-dependent chemical potential as
\begin{equation}
\mu_{\chi}=\big(1+\chi\gamma^{3}\big)^{1/3}\mu,
\label{muchi}
\end{equation}
where we have introduced a dimensionless ratio between the pumped charge and the
equilibrium charge,
\begin{eqnarray}
\gamma=\Big(\frac{3e^{2}\hbar v_{F}^{3}{\bf E}\cdot{\bf B}\tau_{v}}{2\mu^{3}}\Big)^{1/3}.\label{mupm}
\end{eqnarray}
It follows from Eq.~\eqref{mupm} that by tuning the external fields the system undergoes
a chirality-dependent LT at $\gamma=\pm1$, i.e., the change of the topology of the chirality-dependent Fermi surface. In the following we shall work in the weak magnetic
field limit \cite{BNLL}, thus neglecting the Landau level structure of Weyl nodes \cite{Parameswaran,HosurQi,GoswamiOptics,AshbyABJ}. In addition, we will focus on
the $n$-doped case with a finite positive equilibrium chemical potential $\mu>0$
throughout this paper (the discussion of the $p\textnormal{-}$doped case is similar).

\begin{figure}[t]
\centering
\subfigure{\includegraphics[width=6cm]{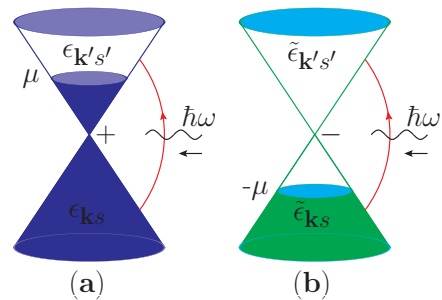}}
\caption{(Color online) The distribution of electrons in the two Weyl nodes $\chi=\pm$
induced by the chiral anomaly (${\bf E}\cdot{\bf B}\neq0$) in undoped Weyl SMs.}
\label{phs}
\end{figure}

It has been demonstrated that no plasmon exists in Dirac SMs \cite{HwangDasSarma}
or undoped Weyl SMs \cite{LvZhang} within the random phase approximation (RPA).
However, when the chiral anomaly occurs, the anomalous charge transfer between
the two Weyl nodes forces the Fermi surfaces to move away from their
equilibrium position in opposite directions as shown in Fig.~\ref{phs}.
Thus the chemical potentials of the two Weyl nodes are $\mu_{+}$
and $\mu_{-}$, satisfying the relation $\mu_{+}=-\mu_{-}\equiv\mu>0$.
In principle, the metallic nature of intrinsic Weyl SMs with chiral
anomaly would support plasmon modes. In the following, we present
an exact and general proof of the existence of the plasmon due
to the chiral anomaly in undoped Weyl SMs.

The general form of the wave vector $q$- and frequency $\omega$-dependent
dielectric function within the RPA is given by
\begin{eqnarray}
\varepsilon(q,\omega)=1-V(q)\Pi(q,\omega),
\end{eqnarray}
where $V(q)=4\pi e^{2}/\kappa q^{2}$ is the Fourier transform of the 3D
Coulomb interaction, with $\kappa$ being the effective dielectric constant.
Let us consider one of the Weyl nodes. The noninteracting polarization function
$\Pi(q,\omega)$ reads~(see Appendix \ref{Sec:AppendixA})
\begin{eqnarray}
\Pi(q,\omega)=\frac{g}{L^{3}}\sum_{{\bf k}ss^{\prime}}
\frac{f(\epsilon_{{\bf k}s})-f(\epsilon_{{\bf k^{\prime}}s^{\prime}})}
{\hbar\omega+\epsilon_{{\bf k}s}-\epsilon_{{\bf k^{\prime}}s^{\prime}}+i\eta}
F_{ss^{\prime}}({\bf k},{\bf k^{\prime}}),\label{DefPol}
\end{eqnarray}
where $g$ is the number of pairs of Weyl nodes, $\eta$ is a positive infinitesimal,
and $s,s^{\prime}=\pm$ are the band indices. The overlap of eigenstates
$F_{ss^{\prime}}({\bf k},{\bf k^{\prime}})$ is given by
\begin{eqnarray}
F_{ss^{\prime}}({\bf k},{\bf k^{\prime}})=\frac{1+ss^{\prime}\cos\theta_{{\bf k}{\bf k^{\prime}}}}{2},\label{Overlap}
\end{eqnarray}
where $\theta_{{\bf k}{\bf k^{\prime}}}$ is the angle between the 3D wave vectors
${\bf k^{\prime}}$ and ${\bf k}$ with ${\bf k}^{\prime}={\bf k+q}$. Here $f\left(x\right)=\left[1+\exp\left\{ \beta(x-\mu)\right\} \right]^{-1}$ is the Fermi
distribution function with $\beta=1/k_{B}T$.

To proceed with the theoretical details, we assume zero temperature $T=0\:\mathrm{K}$.
The Fermi distribution function $f\left(x\right)$ turns into a simple step function $\theta(\mu-x)$. Because of the general relation of the polarization function $\Pi(q,-\omega)=[\Pi(q,\omega)]^{\ast}$, we can restrict our discussion to the
positive frequency case $\omega>0$. In the rest of the calculation, we will set
$\hbar=v_{F}=1$, which immediately implies the relation of $\mu=k_{F}$.

%
%

\section{undoped Weyl semimetals\label{Sec:undoped}}

In general, for a system with particle-hole symmetry, its polarization
function depends only on the magnitude of chemical potential~(see Appendix
\ref{Sec:AppendixB}), namely, $\Pi\left(q,\omega,\mu\right)=\Pi\left(q,\omega,-\mu\right)
=\Pi\left(q,\omega,\left|\mu\right|\right)$. Therefore, we come to a
conclusion that the polarization function of undoped Weyl SMs with the
chirality-dependent chemical potentials $\mu_{\pm}$ is identical to that
of doped Weyl SMs with a chemical potential $\mu=|\mu_\pm|$.

Next we set out to find the plasmon dispersion, which can be obtained within the
RPA by finding the zeros of the dielectric function,
\begin{eqnarray}
\varepsilon(q,\omega-i\Gamma)=0,
\label{plasmon}
\end{eqnarray}
where $\Gamma$ is the decay rate of the plasmon. For weak damping, Eq.~(\ref{plasmon})
reduces to the following approximate equation,
\begin{eqnarray}
\mathrm{Re}~\varepsilon(q,\omega)=0.
\label{approxplasmon}
\end{eqnarray}
For the long wavelength approximation $q\ll\omega\ll\mu$, due to $\mathrm{Im}~\varepsilon(q\to0,\omega)=0$, Eq.~(\ref{plasmon})
reduces to
\begin{eqnarray}
\mathrm{Re}~\varepsilon(q\to0,\omega)=0.
\label{longplasmon}
\end{eqnarray}
To order $q^0$,
the real part of the dielectric function has the form
\begin{eqnarray}
\mathrm{Re}~\varepsilon(q\to0,\omega)=\kappa^{\ast}(\omega)
-\frac{4\alpha_{\kappa}g\mu^2}{3\pi\omega^2},
\label{dieleclong10}
\end{eqnarray}
where the function $\kappa^{\ast}(\omega)$ is defined as
\begin{eqnarray}
\kappa^{\ast}(\omega)=1+\frac{\alpha_{\kappa}g}{3\pi}
\log\Big|\frac{4\Lambda^2}{4\mu^2-\omega^2}\Big|
\label{kappastar1}
\end{eqnarray}
with $\alpha_{\kappa}=e^2/\kappa$ the effective fine structure constant.
Substituting Eq.~(\ref{dieleclong10}) into Eq.~(\ref{longplasmon}) yields
\begin{eqnarray}
\omega_{0}=\sqrt{\frac{\alpha_{\kappa}}{\kappa^{\ast}(\omega_{0})}}
\sqrt{\frac{4g\mu^2}{3\pi}}.
\label{longplasmon10}
\end{eqnarray}
Neglecting the logarithmic corrections in Eq.~(\ref{kappastar1}), we can obtain the lowest
plasmon frequency $\omega_{0}\approx\sqrt{\frac{4g\alpha_{\kappa}\mu^{2}}{3\pi}}$.
Two remarks are in order here. First, the linear dependence of $\omega_{0}$ on
$\mbox{\ensuremath{\mu}}$ also holds for the Dirac SMs \cite{HwangDasSarma} and
Weyl SMs in the absence of the chiral anomaly \cite{LvZhang}. Second, recalling
the chirality-dependent chemical potential in Eq.~(\ref{muchi}), one immediately
finds that $\omega_{0}\propto\left|\mathbf{B}\right|^{1/3}$. Note that this
result is different from previous work in the limit of a strong magnetic field \cite{SonSpivak,panfilov}. In Ref.~\onlinecite{SonSpivak} the plasmon frequency
is found to be proportional to $|\mathbf{B}|^{1/2}$ in the intrinsic case, whereas Ref.~\onlinecite{panfilov} only considered the doped case.

\begin{figure}[htbp]
\centering
\subfigure{\includegraphics[width=6cm]{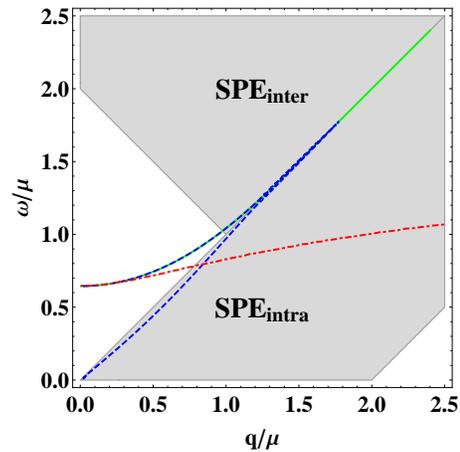}}
\caption{(Color online) Plasmon modes in undoped 3D Weyl SMs with chiral
anomaly, i.e., ${\bf E}\cdot{\bf B}\neq0$ are calculated within the RPA.
The red dotdashed line shows the long wavelength plasmon mode, the blue
dashed line corresponds to the approximate solution obtained from
Eq.~(\ref{approxplasmon}), and the green solid line represents the exact
solution of Eq.~(\ref{plasmon}). The shaded area indicates the intraband
and interband SPE regions\cite{parameters}.}\label{PlasmonWOA}
\end{figure}
Taking into account the leading order contribution of $q$, we get
\begin{eqnarray}
&&\mathrm{Re}~\varepsilon(q\to0,\omega)\label{dieleclong12}\\
&&=\kappa^{\ast}(\omega)-\frac{4\alpha_{\kappa}g\mu^2}{3\pi\omega^2}
\Big[1-\frac{q^2}{(2\mu)^2}\big(1+\mathcal{F}(2\mu,\omega)\big)\Big],\nonumber
\end{eqnarray}
with $\mathcal{F}(x,y)=\frac{x^4(y^2-\frac{3}{5}x^2)}{y^2(x^2-y^2)^2}$. To gain
some insight into the long wavelength plasmon dispersion, we write down an
approximate expression from Eqs.~(\ref{longplasmon}) and (\ref{dieleclong12}) as
\begin{eqnarray}
\omega\approx\omega_0\Big[1-\frac{q^2}{8\mu^2}
\big(1+\mathcal{F}(2\mu,\omega_{0})\big)\Big].
\end{eqnarray}

For comparison, we compute the exact solution, the long wavelength solution, and
the approximate solution, respectively, which are plotted in Fig.~\ref{PlasmonWOA}.
In the long wavelength regime, all the three solutions are in good agreement with
each other~\cite{parameters}. The neglect of the logarithmic term in Eq.~(\ref{kappastar1})
that underlies measurable consequences \cite{BRoy} would enhance the plasmon
frequency. It should be noted that the lower branch of the approximate solution
is fully in the intraband single particle excitation (SPE) region, which is merely
an artifact due to the weak damping approximation of Eq.~(\ref{plasmon}).

\section{doped Weyl semimetals\label{Sec:doped}}
\begin{figure}[t]
\centering
\subfigure{\includegraphics[width=4.0cm]{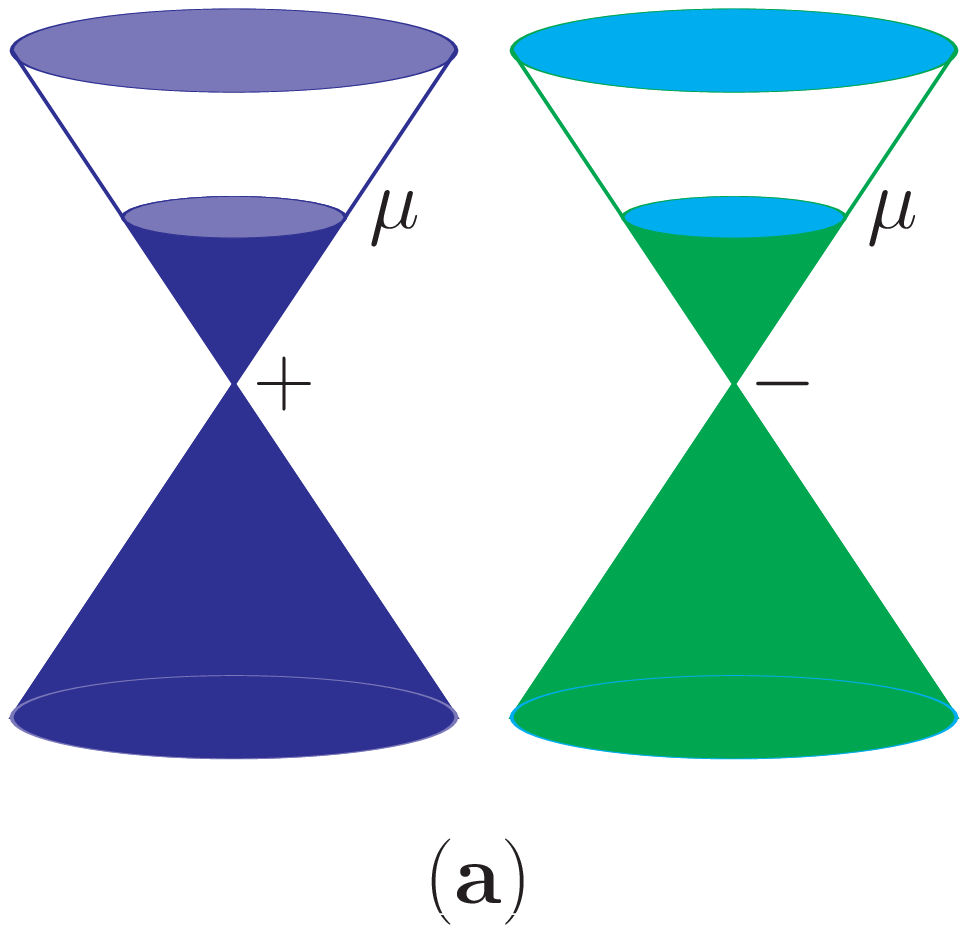}}\hspace{0.5cm}
\subfigure{\includegraphics[width=4.0cm]{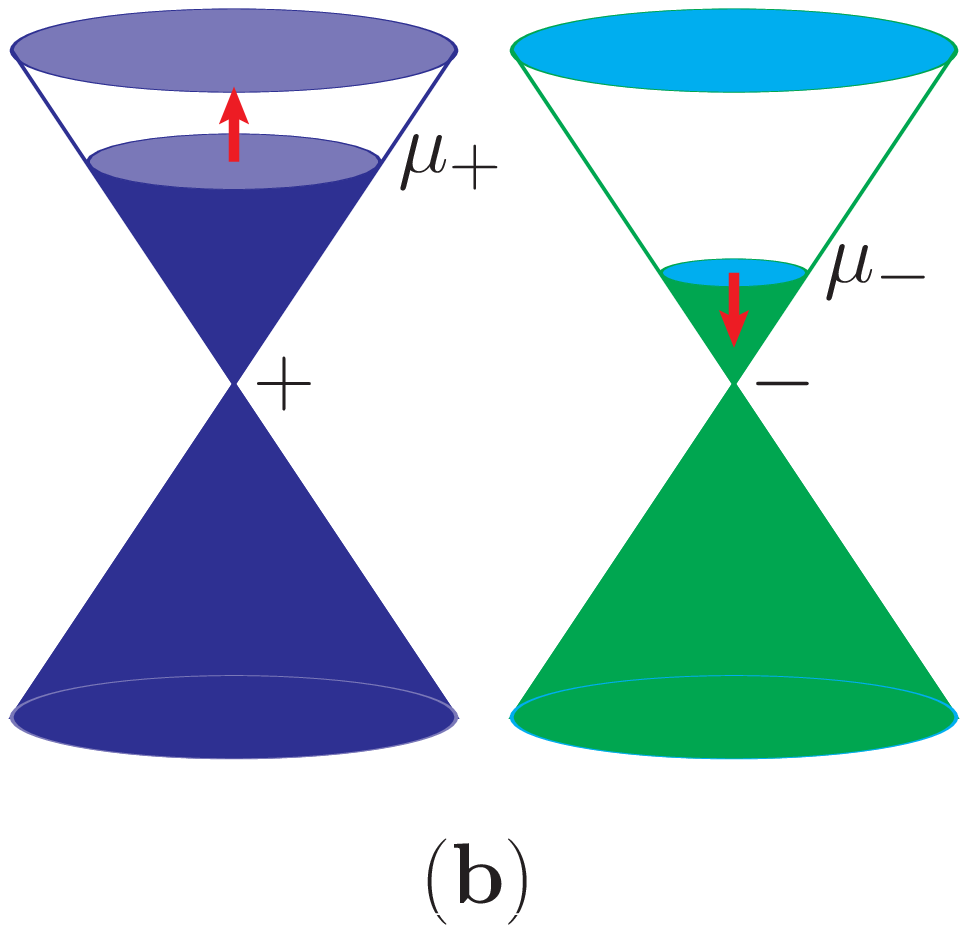}}\\
\caption{(Color online) (a) Equilibrium distribution of electrons in
the two Weyl nodes without chiral anomaly (${\bf E}\cdot{\bf B}=0$).
(b) Fermi levels of the Weyl nodes shift away from the equilibrium position
$\mu$ due to the chiral anomaly (${\bf E}\cdot{\bf B}\neq0$). $\mu_{\pm}$
are the resulting chirality-dependent chemical potentials.}
\label{valleys}
\end{figure}

%
Now we turn to investigate the effect of the chiral anomaly on the plasmon mode in
doped Weyl SMs. Simultaneously turning on the parallel electric field $\mathbf{E}$
and magnetic field $\mathbf{B}$, the amount of electrons transferred from one Weyl
node to the other is equal to $\Delta\rho_{\chi}$, with the result that the Fermi
surface of one Weyl node shifts upward and the other shifts downward (see Fig.~\ref{valleys}).
We assign the chirality-dependent chemical potential $\mu_{\pm}$ for the large and
small Fermi surfaces with $\mu_{+}>|\mu_{-}|$.
In the long wavelength approximation $q\ll\omega\ll|\mu_{-}|\le\mu_{+}$,
to order $q^2$, the real part of the polarization function takes the form
\begin{eqnarray}
&&\mathrm{Re}~\varepsilon(q\to0,\omega)=\kappa^{\ast}(\omega)
-\frac{2\alpha_{\kappa}g(\mu_{+}^2+\mu_{-}^2)}{3\pi\omega^2}\nonumber\\
&&\times\Big[1-\frac{q^2}{4(\mu_{+}^2+\mu_{-}^2)}
\sum_{\lambda=\pm}\big(1+\mathcal{F}(2\mu_{\lambda},\omega_{0})\big)\Big].
\label{dieleclong22}
\end{eqnarray}
To obtain an approximate behavior of the long wavelength plasmon dispersion, one can
arrive at an expression from Eqs.~(\ref{longplasmon}) and (\ref{dieleclong22}) as
\begin{eqnarray}
\omega\approx\omega_{0}\Big[1-\frac{q^2}{8(\mu_{+}^2+\mu_{-}^2)}\sum_{\lambda=\pm}
\big(1+\mathcal{F}(2\mu_{\lambda},\omega_{0})\big)\Big].
\end{eqnarray}
where the notations $\kappa^{\ast}(\omega)$ and $\omega_{0}$ are given by
\begin{eqnarray}
&&\omega_{0}=\sqrt{\frac{\alpha_{\kappa}}{\kappa^{\ast}(\omega_{0})}}
\sqrt{\frac{2g}{3\pi}(\mu_{+}^{2}+\mu_{-}^{2})},
\label{longplasmon20}\\
&&\kappa^{\ast}(\omega)=1+\frac{\alpha_{\kappa}g}{6\pi}
\left(\sum_{\lambda=\pm}\log\Big|\frac{4\Lambda^2}{4\mu_{\lambda}^{2}-\omega^2}\Big|\right),
\label{kappastar2}
\end{eqnarray}
which can be traced back to the counterpart of Eq.~(\ref{longplasmon10}) in the
undoped case by taking $\mu_{+}=\mu_{-}$.

\begin{figure}[t]
\centering
\subfigure{\includegraphics[width=4.0cm]{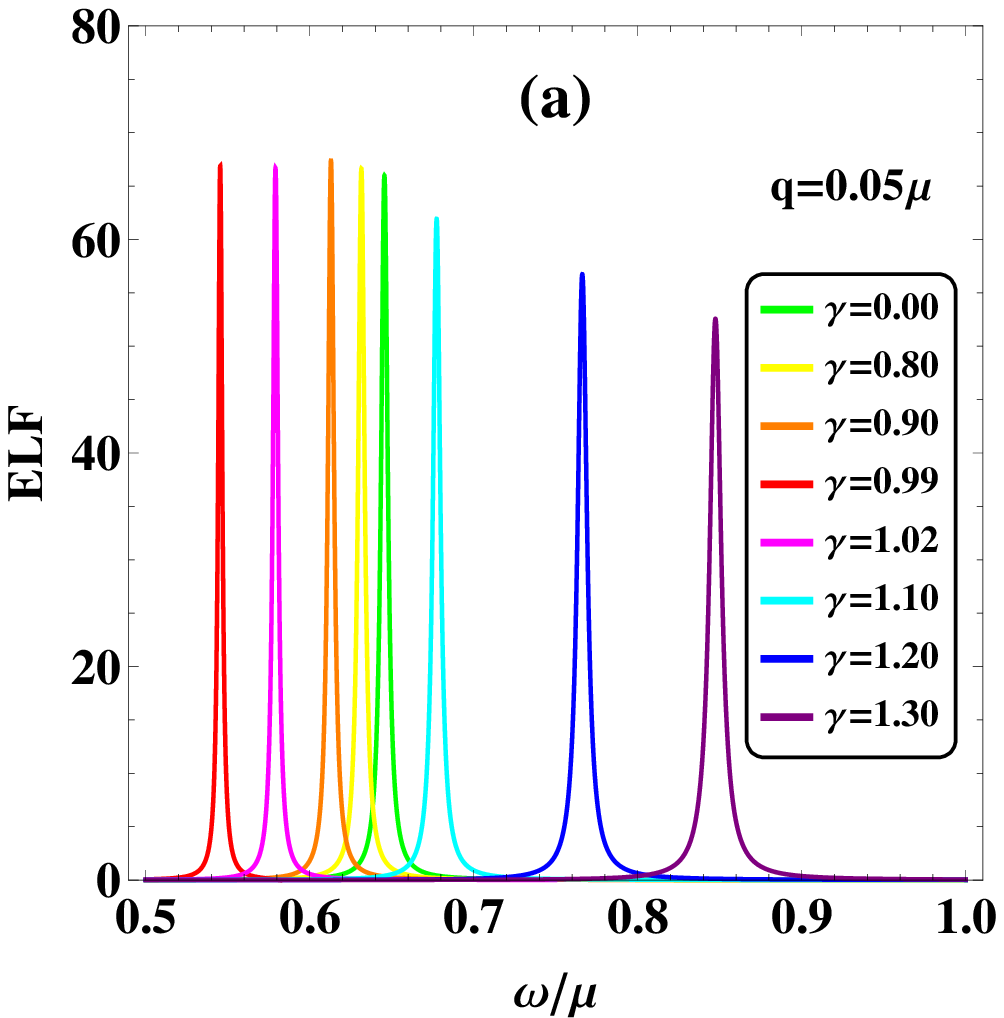}}
\subfigure{\includegraphics[width=4.0cm]{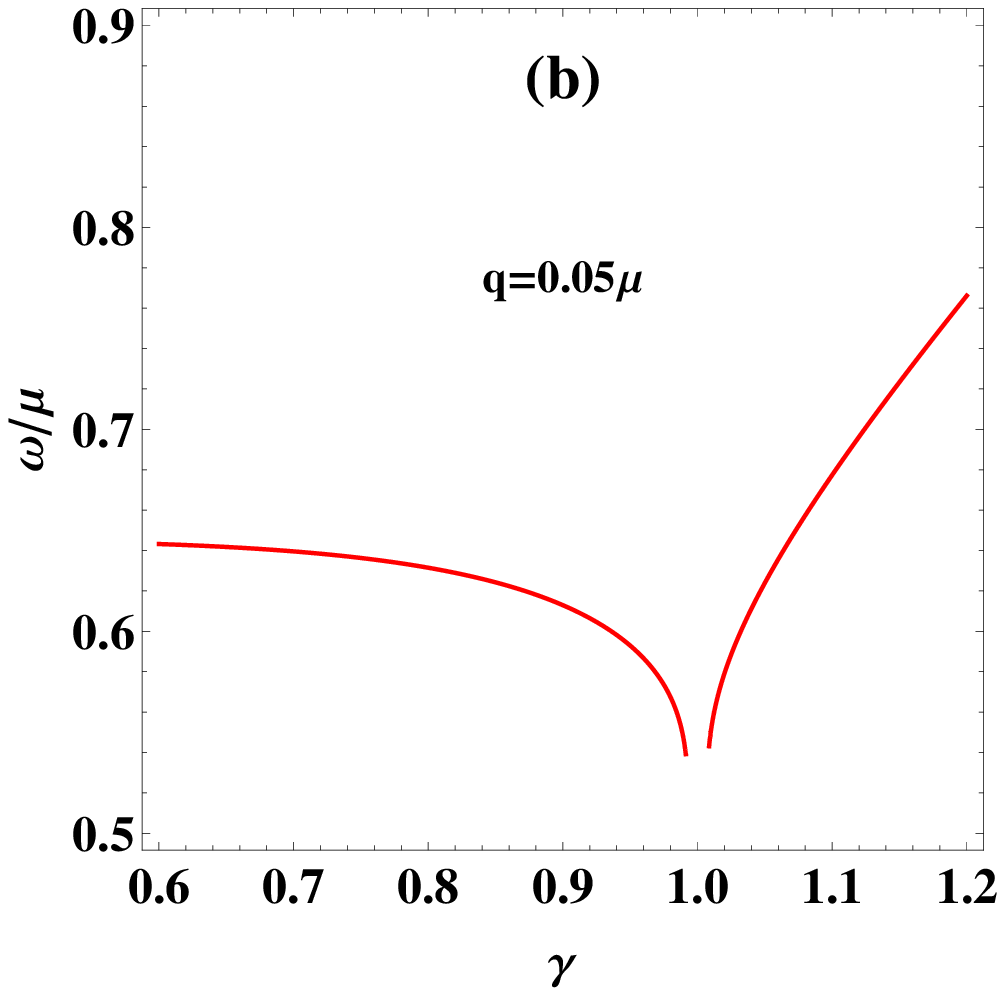}}\\
\caption{(Color online) (a) Energy-loss function for a series of
values of $\gamma$ in doped Weyl SMs with the chiral anomaly.
(b) Dependence of frequencies of the undamped plasmon mode on the ratio $\gamma$.
The LT occurs at the turning point of the plasmon dispersion \cite{parameters}.}
\label{PlasmonELF}
\end{figure}

The plasmons can also be revealed as sharp peaks in the energy loss function
(ELF), defined as the imaginary part of the inverse dielectric function, i.e., $\mathrm{Im}\left[1/\varepsilon(q,\omega)\right]$, that can be probed in various
spectroscopy experiments, such as the electron energy-loss spectroscopy. As shown
in Fig.~\ref{PlasmonELF}(a), in the presence of the chiral anomaly the plasmon
exhibits some exotic features in the ELF spectrum. As long as the ratio $\gamma$
gradually increases from $0$ to $1$, the plasmon frequency with chiral anomaly
$\omega_{\mathrm{ch}}$ has a redshift with respect to the frequency without chiral
anomaly $\omega_{\mathrm{eq}}$ and finally reaches a minimum $\omega_{\mathrm{min}}$.
On the other hand, once the small Fermi surface crosses the Weyl node, i.e.,
$\gamma>1$, the plasmon frequency becomes larger than $\omega_{\mathrm{min}}$ and
then has a continuous violetshift. The behavior of plasmons of doped Weyl SMs under
the influence of the chiral anomaly can be captured by our long wavelength expressions
in Eq.~(\ref{longplasmon20}) and summarized as follows
\begin{eqnarray}
\left\{
\begin{array}{ll}
\displaystyle \omega_{\mathrm{min}}<\omega_{\mathrm{ch}}<\omega_{\mathrm{eq}}, & {\rm 0<\gamma<1,}\medskip\\
\displaystyle \omega_{\mathrm{min}}<\omega_{\mathrm{ch}}, & {\rm 1<\gamma.}\\
\end{array}
\right.
\end{eqnarray}
Therefore, the unique features of undamped plasmon mode can clearly characterize
the chiral anomaly in doped Weyl SMs. It should be emphasized that compared with
other methods\cite{Parameswaran,HosurQi,GoswamiOptics,AshbyABJ}, our method possesses
the advantage that we can directly determine the position of the chirality-dependent
LT point from the plasmon dispersion, as shown in Fig.~\ref{PlasmonELF}(b), which
coincides with the minimal frequency of plasmon mode. Actually, when $|\gamma-1|\to0$,
the small Fermi level is close to the Weyl node point, such that small energy or
momentum could induce interband transition and lead to a large number of electron-hole excitations. The plasmon mode will be damped by these electron-hole excitations. Increasing
$q$ will broaden the damping region [see Fig.~\ref{PlasmonELF}(b)]. Hence, the fate of
the plasmon mode indeed connects with the chirality-dependent LT.

\section{CONCLUSIONS\label{Sec:CONCLUSIONS}}

In summary, we investigated the chiral anomaly effect on the plasmon mode in 3D
Weyl SMs within the RPA. We proved that a different plasmon mode would emerge in undoped
Weyl SMs due to the chiral anomaly. We also demonstrated the unusual properties
of the plasmon mode in doped Weyl SMs and further pointed out that the plasmon can
be taken as a fingerprint of the chiral anomaly. Finally, we showed how to identify
the chirality-dependent LT point from the plasmon dispersion. Our work sheds light
on the probing of the chiral anomaly in 3D Weyl SMs via the plasmon mode. The tunability
of plasmons due to the chiral anomaly also makes Weyl SMs promising candidates for
plasmonics \cite{Maier}.

\section*{ACKNOWLEDGMENTS\label{Sec:acknowledgments}}

We are grateful to Yuanpei Lan, Min Lv, Wen-Yu Shan, and Hui Zhang for stimulating
discussions, and to Ran Cheng, Matthew Daniels, and Furu Zhang for a careful reading of
the manuscript. This work is supported by the National Science Foundation (EFRI-1433496),
the Air Force Office of Scientific Research (FA9550-14-1-0277), the Natural Science
Foundation of Sichuan Educational Committee(Grant No. 13ZB0157), the Natural Science
Foundation of Sichuan Normal University(Grant No. 15YB001), and the National Natural
Science Foundation of China (Grant No. 11274286)

\appendix

\begin{widetext}

\section{The calculation of the polarization function with positive chemical potential\label{Sec:AppendixA}}

In this Appendix we present the major steps of calculating
the polarization function of a Weyl node with positive chemical potential $+\mu$ in the 3D Weyl semimetal.
Those of the other Weyl nodes can be obtained in a similar manner.
The polarization function can be decomposed into two parts,
\begin{eqnarray}
\Pi(q,\omega)=\Pi^{-}(q,\omega)+\Pi^{+}(q,\omega),\label{Poltot}
\end{eqnarray}
where $\Pi^{\pm}(q,\omega)$ are defined by
\begin{eqnarray}
\Pi^{-}(q,\omega)=\frac{g}{L^3}\sum_{{\bf k}}\left(
\frac{[f(\epsilon_{{\bf k}-})-f(\epsilon_{{\bf k^{\prime}}-})]
(1+\cos\theta_{{\bf k}{\bf k^{\prime}}})/2}
{\hbar\omega+\epsilon_{{\bf k}-}-\epsilon_{{\bf k^{\prime}}-}+i\eta}
+\frac{f(\epsilon_{{\bf k}-})(1-\cos\theta_{{\bf k}{\bf k^{\prime}}})/2}
{\hbar\omega+\epsilon_{{\bf k}-}-\epsilon_{{\bf k^{\prime}}+}+i\eta}
-\frac{f(\epsilon_{{\bf k^{\prime}}-})(1-\cos\theta_{{\bf k}{\bf k^{\prime}}})/2}
{\hbar\omega+\epsilon_{{\bf k}+}-\epsilon_{{\bf k^{\prime}}-}+i\eta}
\right),\\
\Pi^{+}(q,\omega)=\frac{g}{L^3}\sum_{{\bf k}}\left(
\frac{[f(\epsilon_{{\bf k}+})-f(\epsilon_{{\bf k^{\prime}}+})]
(1+\cos\theta_{{\bf k}{\bf k^{\prime}}})/2}
{\hbar\omega+\epsilon_{{\bf k}+}-\epsilon_{{\bf k^{\prime}}+}+i\eta}
+\frac{f(\epsilon_{{\bf k}+})(1-\cos\theta_{{\bf k}{\bf k^{\prime}}})/2}
{\hbar\omega+\epsilon_{{\bf k}+}-\epsilon_{{\bf k^{\prime}}-}+i\eta}
-\frac{f(\epsilon_{{\bf k^{\prime}}+})(1-\cos\theta_{{\bf k}{\bf k^{\prime}}})/2}
{\hbar\omega+\epsilon_{{\bf k}-}-\epsilon_{{\bf k^{\prime}}+}+i\eta}
\right).
\end{eqnarray}

Due to the causality $\mathrm{Re}~\Pi^{-}(q,-\omega)=\mathrm{Re}~\Pi^{-}(q,\omega)$,
in the following we focus only on the case for $\omega>0$. We first evaluate the
polarization function of the intrinsic case with $\mu=0$ that implies $\Pi^{+}(q,\omega)$ vanishes. After some simple algebra, we can obtain
\begin{eqnarray}
\Pi^{-}(q,\omega)=-\frac{g}{16\pi^2q}\int_{0}^{\Lambda}dk
\int_{|k-q|}^{k+q}dk^{\prime}\left[(k^{\prime}-k)^2-q^2\right]
\left(\frac{1}{\omega-k-k^{\prime}+i\eta}
-\frac{1}{\omega+k+k^{\prime}+i\eta}\right),
\end{eqnarray}
where $\Lambda$ is the cutoff. Using the Dirac identity
$\frac{1}{x\pm i\eta}=\mathcal{P}\frac{1}{x}\mp i\pi\delta(x)$
one can get
\begin{eqnarray}
&&\mathrm{Im}~\Pi^{-}(q,\omega)=\frac{g}{16\pi q}\int_{0}^{\Lambda}
dk\int_{|k-q|}^{k+q}dk^{\prime}\left[(k^{\prime}-k)^2-q^2\right]
\delta(\omega-k-k^{\prime}),
\label{PolintIm0}\\
&&\mathrm{Re}~\Pi^{-}(q,\omega)=-\frac{g}{16\pi^2q}\mathcal{P}
\int_{0}^{\Lambda}dk\int_{|k-q|}^{k+q}dk^{\prime}
\left[(k^{\prime}-k)^2-q^2\right]\left(\frac{1}{\omega-k-k^{\prime}}
-\frac{1}{\omega+k+k^{\prime}}\right),\label{PolintRe0}
\end{eqnarray}
where the notation $\mathcal{P}$ means the principal value of the integral.
It is straightforward to calculate the imaginary part of the intrinsic
polarization function
\begin{eqnarray}
\mathrm{Im}~\Pi^{-}(q,\omega)=-\frac{gq^2\theta(\omega-q)}{24\pi}.
\label{PolintIm}
\end{eqnarray}

In fact, there are two different methods to calculate the real part of the
polarization function. One is to directly carry out the integral. The other
is to apply the Kramers-Kr\"{o}nig relation.

We at first perform the integration Eq. (\ref{PolintRe0}) to get the real
part of the intrinsic polarization function. It is convenient to decompose
this real part into two terms,
\begin{eqnarray}
\mathrm{Re}~\Pi^{-}(q,\omega)=\mathrm{Re}~\Pi_{1}^{-}(q,\omega)
+\mathrm{Re}~\Pi_{2}^{-}(q,\omega),
\label{RePi}
\end{eqnarray}
where
\begin{eqnarray}
&&\mathrm{Re}~\Pi_{1}^{-}(q,\omega)=\frac{g}{16\pi^2q}\mathcal{P}
\int_{0}^{\Lambda}dk\int_{|k-q|}^{k+q}dk^{\prime}
\left((-3k+k^{\prime})+\frac{(2k+\omega)^2-q^2}{k^{\prime}+k+\omega}\right),\\
&&\mathrm{Re}~\Pi_{2}^{-}(q,\omega)=\frac{g}{16\pi^2q}\mathcal{P}
\int_{0}^{\Lambda}dk\int_{|k-q|}^{k+q}dk^{\prime}
\left((-3k+k^{\prime})+\frac{(2k-\omega)^2-q^2}{k^{\prime}+k-\omega}\right).
\end{eqnarray}
After some cumbersome but straightforward calculation, we can get
\begin{eqnarray}
&&\mathrm{Re}~\Pi_{1}^{-}(q,\omega)
=-\frac{2gq^2}{96\pi^2}\log\frac{(2\Lambda+\omega)^2-q^2}{(q+\omega)^2}
+\frac{g}{96\pi^2}\left[+6\omega(q+2\Lambda)\right]\nonumber\\
&&+\frac{g}{96\pi^2}\left[\frac{(2\Lambda+\omega)^3}{q}
\log\frac{2\Lambda+\omega+q}{2\Lambda+\omega-q}
-2(2\Lambda+\omega)^2-3q(2\Lambda+\omega)\log\frac{2\Lambda+\omega+q}
{2\Lambda+\omega-q}+\frac{16}{3}q^2\right],\\
&&\mathrm{Re}~\Pi_{2}^{-}(q,\omega)
=-\frac{2gq^2}{96\pi^2}\log\frac{(2\Lambda-\omega)^2-q^2}{(q-\omega)^2}
+\frac{g}{96\pi^2}\left[-6\omega(q+2\Lambda)\right]\nonumber\\
&&+\frac{g}{96\pi^2}\left[\frac{(2\Lambda-\omega)^3}{q}
\log\frac{2\Lambda-\omega+q}{2\Lambda-\omega-q}
-2(2\Lambda-\omega)^2-3q(2\Lambda-\omega)\log\frac{2\Lambda-\omega+q}
{2\Lambda-\omega-q}+\frac{16}{3}q^2\right].
\end{eqnarray}
and
\begin{eqnarray}
&&\mathrm{Re}~\Pi^{-}(q,\omega)=
-\frac{2gq^2}{96\pi^2}\left(\log\frac{(2\Lambda+\omega)^2-q^2}{(q+\omega)^2}
+\log\frac{(2\Lambda-\omega)^2-q^2}{(q-\omega)^2}\right)\nonumber\\
&&+\underline{\frac{g}{96\pi^2}\left[\frac{(2\Lambda+\omega)^3}{q}
\log\frac{2\Lambda+\omega+q}{2\Lambda+\omega-q}
-2(2\Lambda+\omega)^2-3q(2\Lambda+\omega)\log\frac{2\Lambda+\omega+q}
{2\Lambda+\omega-q}+\frac{16}{3}q^2\right]}\nonumber\\
&&+\underline{\frac{g}{96\pi^2}\left[\frac{(2\Lambda-\omega)^3}{q}
\log\frac{2\Lambda-\omega+q}{2\Lambda-\omega-q}
-2(2\Lambda-\omega)^2-3q(2\Lambda-\omega)\log\frac{2\Lambda-\omega+q}
{2\Lambda-\omega-q}+\frac{16}{3}q^2\right]}.
\label{PolextRet}
\end{eqnarray}
Two remarks about the real part of the intrinsic polarization function are in order
here. First, $\mathrm{Re}~\Pi_{2}^{-}(q,\omega)$ can also be obtained by replacing
$\omega$ with $-\omega$ in $\mathrm{Re}~\Pi_{1}^{-}(q,\omega)$. Second, it can be
seen that $\mathrm{Re}~\Pi^{-}(q,\omega)$ is an even function in $\omega$ and will
be valid for an arbitrary frequency.

Since the cutoff $\Lambda$ is much larger than both $q$ and $\omega$, it is
instructive to take a look at the expression of $\mathrm{Re}~\Pi^{-}(q,\omega)$
in large $\Lambda$ limit. Making use of the limits
\begin{eqnarray}
\lim_{\frac{t}{q}\to\infty}\left(\frac{t^3}{q}\log\frac{t+q}{t-q}-2t^2\right)
=\frac{2q^2}{3},\hspace{1cm}
\lim_{\frac{t}{q}\to\infty}t \log\frac{t+q}{t-q}=2q,
\end{eqnarray}
we immediately verify that these underlined terms in Eq. (\ref{PolextRet}) vanish with $t=2\Lambda\pm\omega$ and then get a simple expression of $\mathrm{Re}~\Pi^{-}(q,\omega)$
\begin{eqnarray}
\mathrm{Re}~\Pi^{-}(q,\omega)=-\frac{gq^2}{48\pi^2}
\log\left(\frac{(2\Lambda+\omega)^2-q^2}{(q+\omega)^2}
\frac{(2\Lambda-\omega)^2-q^2}{(q-\omega)^2}\right).
\label{PolintRefinal1}
\end{eqnarray}
We can further simplify the terms of $q$ and $\omega$ in the numerator of the
logarithmic function and have
\begin{eqnarray}
\mathrm{Re}~\Pi^{-}(q,\omega)=-\frac{gq^2}{24\pi^2}
\log\Big|\frac{4\Lambda^2}{q^2-\omega^2}\Big|.
\label{PolintRefinal2}
\end{eqnarray}

Now we turn to calculate the real part of the intrinsic polarization function
$\Pi^{-}(q,\omega)$ using the Kramers-Kr\"{o}nig relations. Since the imaginary
part does not approach zero as $\omega\to\infty$, one needs to utilize the generalized
Kramers-Kr\"{o}nig relation with one subtraction\cite{Bjorken},
\begin{eqnarray}
&&\mathrm{Re}~\Pi^{-}(q,\omega)=\mathrm{Re}~\Pi^{-}(q,0)
+\frac{\omega}{\pi}\mathcal{P}\int_{-\infty}^{\infty}
d\xi\frac{\mathrm{Im}~\Pi^{-}(q,\xi)}{\xi(\xi-\omega)},\label{PolintRe1}\\
&&\mathrm{Im}~\Pi^{-}(q,\omega)=\mathrm{Im}~\Pi^{-}(q,0)
-\frac{\omega}{\pi}\mathcal{P}\int_{-\infty}^{\infty}
d\xi\frac{\mathrm{Re}~\Pi^{-}(q,\xi)}{\xi(\xi-\omega)}.\label{PolintIm1}
\end{eqnarray}
The zero frequency term $\mathrm{Re}~\Pi^{-}(q,0)$ in Eq. (\ref{PolintRe1}) can be
obtained from Eq. (\ref{PolintRe0}),
\begin{eqnarray}
&&\mathrm{Re}~\Pi^{-}(q,0)
=-\frac{4gq^2}{96\pi^2}\log\frac{(2\Lambda)^2-q^2}{q^2}
+\underline{\frac{2g}{96\pi^2}\left[\frac{(2\Lambda)^3}{q}
\log\frac{2\Lambda+q}{2\Lambda-q}-2(2\Lambda)^2-3q(2\Lambda)
\log\frac{2\Lambda+q}{2\Lambda-q}+\frac{16}{3}q^2\right]}.
\end{eqnarray}
In the limit of large cutoff $\Lambda$, we find that the above underlined term
vanishes. Neglecting the $q^2$ term in the numerator of the logarithmic function
yields
\begin{eqnarray}
\mathrm{Re}~\Pi^{-}(q,0)=-\frac{gq^2}{24\pi^2}\log\frac{4\Lambda^2}{q^2}.
\label{PolintReq0}
\end{eqnarray}
The second term in Eq.(\ref{PolintRe1}) is calculated by carrying out the integration
\begin{eqnarray}
\mathrm{Re}~\Pi^{-}(q,\omega)-\mathrm{Re}~\Pi^{-}(q,0)=
\frac{\omega}{\pi}\mathcal{P}\int_{-\infty}^{\infty}
d\xi\frac{\mathrm{Im}~\Pi^{-}(q,\xi)}{\xi(\xi-\omega)}
=-\frac{gq^2}{24\pi^2}\log\Big|\frac{q^2}{q^2-\omega^2}\Big|.
\label{PolintReq1}
\end{eqnarray}
Substituting Eq. (\ref{PolintReq0}) into Eq. (\ref{PolintReq1}) leads to the same
result as Eq. (\ref{PolintRefinal2}), which differs slightly from the counterpart
in Ref. \cite{LvZhang}. It should be pointed out that $\mathrm{Re}~\Pi^{-}(q,\omega)$
satisfies Eq. (\ref{PolintIm1}) by considering $\mathrm{Im}~\Pi^{-}(q,0)=0$. Therefore,
the polarization function of the intrinsic case turns out to be
\begin{eqnarray}
\Pi^{-}(q,\omega)=-\frac{gq^2}{24\pi^2}\left[
\log\Big|\frac{4\Lambda^2}{q^2-\omega^2}\Big|+i\pi\theta(\omega-q)\right].
\end{eqnarray}

Following the similar procedure, one can reach the polarization function
$\Pi^{+}(q,\omega)$ of the extrinsic case with $\mu>0$,
\begin{eqnarray}
&&\mathrm{Im}~\Pi^{+}(q,\omega)=-\frac{gq^2}{8\pi^2}
\Big[\theta(q-\omega)\Big(\frac{\pi G(q,\omega)}{q^2}\theta(2\mu+\omega-q)
-\frac{\pi G(q,-\omega)}{q^2}\theta(2\mu-\omega-q)\Big)\nonumber\\
&&\hspace{3.1cm}+\theta(\omega-q)\Big(-\frac{\pi}{3}\theta(2\mu-\omega-q)-\frac{\pi
G(-q,-\omega)}{q^2}\theta(q+\omega-2\mu)\theta(2\mu+q-\omega)\Big)\Big],
\label{PolextIm}\\
&&\mathrm{Re}~\Pi^{+}(q,\omega)=-\frac{gq^2}{8\pi^2}\Big[\frac{8\mu^2}{3q^2}
-\frac{G(q,\omega)H(q,\omega)}{q^2}-\frac{G(-q,\omega)H(-q,\omega)}{q^2}
-\frac{G(q,-\omega)H(q,-\omega)}{q^2}\nonumber\\
&&\hspace{3.1cm}-\frac{G(-q,-\omega)H(-q,-\omega)}{q^2}\Big],
\label{PolextRe}
\end{eqnarray}
where the functions $G(q,\omega)$ and $H(q,\omega)$ are defined by
\begin{eqnarray}
&&G(q,\omega)=\frac{1}{12q}\left[(2\mu+\omega)^3
-3q^2(2\mu+\omega)+2q^3\right],\\
&&H(q,\omega)=\log\Big|\frac{2\mu+\omega-q}
{q-\omega}\Big|.
\end{eqnarray}
Combining $\Pi^{-}(q,\omega)$ with $\Pi^{+}(q,\omega)$, we finally obtain the
total polarization function for a Weyl node with positive chemical potential
$+\mu$ in the 3D Weyl semimetal in Eq. (\ref{Poltot}).
%

\section{The equivalence of the polarization functions with opposite
chemical potentials\label{Sec:AppendixB}}
%

In this appendix we will prove the equivalence of the polarization functions with
opposite chemical potentials for a system with particle-hole symmetry.
The polarization function for the other node with negative chemical potential $-\mu$ can be written as
\begin{eqnarray}
\tilde{\Pi}\left(q,\omega\right)=\frac{g}{L^{3}}\sum_{{\bf k}ss^{\prime}}\frac{\tilde{f}(\tilde{\epsilon}_{{\bf k}s})-\tilde{f}(\tilde{\epsilon}_{{\bf k^{\prime}}s^{\prime}})}{\hbar\omega+\tilde{\epsilon}_{{\bf k}s}-\tilde{\epsilon}_{{\bf k^{\prime}}s^{\prime}}+i\eta}F_{ss^{\prime}}\left({\bf k},{\bf k^{\prime}}\right),\label{Pi21}
\end{eqnarray}
where the function $\tilde{f}(x)$ is defined as $\tilde{f}\left(x\right)=\left[1+\exp\left\{ \beta(x+\mu)\right\} \right]^{-1}$.
The particle-hole symmetry of Weyl nodes enables us to relabel the
energy dispersions in the following way: $\tilde{\epsilon}_{\mathbf{k}s}\rightarrow-\epsilon_{\mathbf{k}s}$
and write
\begin{eqnarray}
\tilde{\Pi}\left(q,\omega\right)=\frac{g}{L^{3}}\sum_{{\bf k}ss^{\prime}}\frac{\tilde{f}(-\epsilon_{{\bf k}s})-\tilde{f}(-\epsilon_{{\bf k^{\prime}}s^{\prime}})}{\hbar\omega-\epsilon_{{\bf k}s}+\epsilon_{{\bf k^{\prime}}s^{\prime}}+i\eta}F_{ss^{\prime}}\left({\bf k},{\bf k^{\prime}}\right).
\end{eqnarray}
Utilizing the property of the Fermi distribution function $f\left(x\right)+\tilde{f}\left(-x\right)=1$ leads to
\begin{eqnarray}
\tilde{\Pi}\left(q,\omega\right)=\frac{g}{L^{3}}\sum_{{\bf k}ss^{\prime}}\frac{f\left(\epsilon_{{\bf k^{\prime}}s^{\prime}}\right)-f\left(\epsilon_{{\bf k}s}\right)}{\hbar\omega+\epsilon_{{\bf k^{\prime}}s^{\prime}}-\epsilon_{{\bf k}s}+i\eta}F_{ss^{\prime}}\left({\bf k},{\bf k^{\prime}}\right).
\end{eqnarray}
One can immediately observe the relation
\begin{eqnarray}
\left[\tilde{\Pi}\left(q,-\omega\right)\right]^{\ast}=\frac{g}{L^{3}}\sum_{{\bf k}ss^{\prime}}\frac{f\left(\epsilon_{{\bf k}s}\right)-f\left(\epsilon_{{\bf k^{\prime}}s^{\prime}}\right)}{\hbar\omega+\epsilon_{{\bf k}s}-\epsilon_{{\bf k^{\prime}}s^{\prime}}+i\eta}F_{ss^{\prime}}\left({\bf k},{\bf k^{\prime}}\right).
\end{eqnarray}
It is obvious that $[\tilde{\Pi}\left(q,-\omega\right)]^{\ast}$
is nothing but the definition of the polarization function with a
positive chemical potential $\mu$. Recalling
the general property of $\Pi\left(q,-\omega\right)=\left[\Pi\left(q,\omega\right)\right]^{\ast}$,
we arrive at the desirable result
\begin{eqnarray}
\Pi(q,\omega)=\tilde{\Pi}(q,\omega).\label{equiv}
\end{eqnarray}

\end{widetext}



\begin{thebibliography}{References}
\bibitem{BNLL} In general, when the energy difference between
the zeroth and the first Landau Levels (LLs)
 $E_{L}=\hbar \omega_{c}$ ($\omega_{c}$ is the cyclotron frequency)
is smaller than the thermal excitation energy $E_{T}=k_{B}T$ or the inverse scattering
time, these broadening effects could smear the LLs. As a result, we can neglect the LL quantization.
\end{thebibliography}
\end{document}